\documentclass[aip, reprint]{revtex4-1}
\usepackage{graphicx}
\usepackage{color}
\usepackage{amssymb,amsmath,amsfonts}
\usepackage[justification=centerlast, singlelinecheck=false, format=plain]{caption}
\usepackage[justification=centerlast, singlelinecheck=false, format=plain]{subcaption}
\usepackage{hyperref}
\usepackage{hyperref}
\hypersetup{
    colorlinks,
    citecolor=blue,
    filecolor=blue,
    linkcolor=blue,
    urlcolor=blue
}

\begin{document}

\title{Topological Chaos in a Three-Dimensional Spherical Fluid Vortex}

\author{Spencer A. Smith*}
\affiliation{	School of Natural Sciences, University of California Merced, Merced, CA 95343}
\affiliation{	Mount Holyoke College, South Hadley, MA 01075}
\author{Joshua Arenson}
\affiliation{	School of Natural Sciences, University of California Merced, Merced, CA 95343}
\author{Eric Roberts}
\affiliation{	School of Natural Sciences, University of California Merced, Merced, CA 95343}
\author{Suzanne Sindi}
\affiliation{	School of Natural Sciences, University of California Merced, Merced, CA 95343}
\author{Kevin A. Mitchell}
\affiliation{	School of Natural Sciences, University of California Merced, Merced, CA 95343}

\date{\today}
\begin{abstract}

In chaotic deterministic systems, seemingly stochastic behavior is generated by relatively simple, though hidden, organizing rules and structures.  Prominent among the tools used to characterize this complexity in 1D and 2D systems are techniques which exploit the topology of dynamically invariant structures.  However, the path to extending many such topological techniques to three dimensions is filled with roadblocks that prevent their application to a wider variety of physical systems.  Here, we overcome these roadblocks and successfully analyze a realistic model of 3D fluid advection, by extending the homotopic lobe dynamics (HLD) technique, previously developed for 2D area-preserving dynamics, to 3D volume-preserving dynamics.  We start with numerically-generated finite-time chaotic-scattering data for particles entrained in a spherical fluid vortex, and use this data to build a symbolic representation of the dynamics.  We then use this symbolic representation to explain and predict the self-similar fractal structure of the scattering data, to compute bounds on the topological entropy, a fundamental measure of mixing, and to discover two different mixing mechanisms, which stretch 2D material surfaces and 1D material curves in distinct ways.
 
\end{abstract}

\maketitle

	
The essential allure of chaotic dynamics is confronting a complex, seemingly random, physical process and discovering the hidden, underlying patterns that order it.  This is typified by the seminal experiments of Gollub and Swinney, showing that the progression from regular to turbulent fluid flow occurs via the predictable sequence of period doubling cascades.\cite{GollubSwinney}  Immense success has been achieved in unraveling such patterns for chaotic systems reducible to maps on a one- or two-dimensional phase space\cite{Li75,Sharkovskii64,Milnor88, Thurston88, Easton86}. This has been achieved through a deep understanding of the shape of geometric objects living within the dynamical phase space, using topological tools such as Markov partitions\cite{Guckenheimer83}, symbolic dynamics\cite{LindMarcus, Kitchens, Gilmore11}, braid theory\cite{Boyland00,Thiffeault06,Thiffeault10,Budisic15,Stremler11}, and mapping class groups.\cite{Thurston88,Boyland94,Bestvina95}  A key theme in such studies is topological ÒforcingÓ: the existence of certain short-time structures (e.g. low-period orbits) forces the existence of infinitely many longer-time structures.  The resulting patterns are typically fractal, with symbolic rules describing a rich self-similarity.  Thus, early-time, low-resolution data predicts long-time, high-resolution patterns. This is nicely illustrated by the famous period-three-implies-chaos result: the existence of a single period-three orbit of a map on the unit interval guarantees the existence of periodic orbits of arbitrary period\cite{Li75,Sharkovskii64}.  A central challenge in dynamical systems is extending these topological techniques to higher dimensions, for which there have been few clear paths forward\cite{Lefranc06,Lefranc08, Jung10,Drotos14,Drotos16}.  We demonstrate for the first time how chaotic scattering data for an explicit \emph{three}-dimensional system, a numerical model of a chaotic time-periodic fluid vortex, reveals fractal rules of the dynamics.

\begin{figure*}[htbp]
		\includegraphics[width = \textwidth]{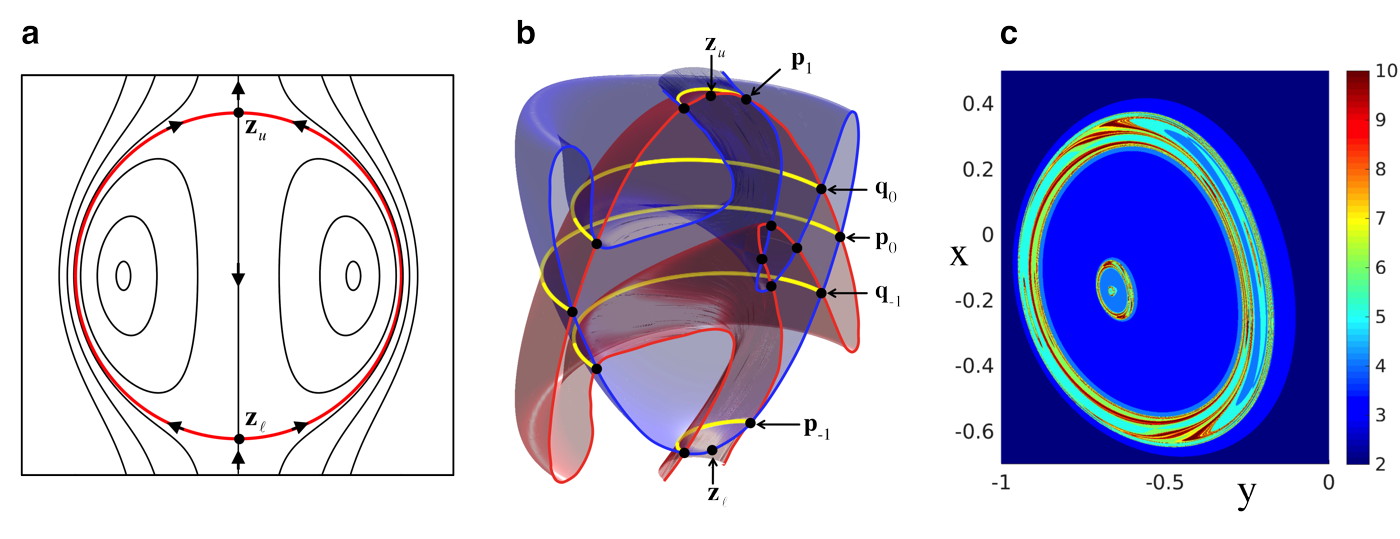}
		\caption{\textbf{Hill's Vortex, Original and Modified.} \textbf{a}, Cross-sectional view of Hill's spherical vortex, including streamlines, the unstable fixed points ($\mathbf{z}_u$ and $\mathbf{z}_{\ell}$), the 2D separatrix (degenerate 2D stable manifold, $W^S$, and unstable manifold, $W^U$, of $\mathbf{z}_u$ and $\mathbf{z}_{\ell}$ respectively, in red), and the degenerate 1D stable and unstable manifolds along the vertical axis.  \textbf{b}, 3D cut-away view of the modified Hill's vortex.  The 2D $W^U$ of $\mathbf{z}_{\ell}$ (blue) and 2D $W^S$ of $\mathbf{z}_u$ (red) are now distinct, forming the heteroclinic tangle.  The primary intersection curve, $\mathbf{p}_0$, which defines the unstable cap, $W^U\left[\mathbf{p}_0 \right]$, and stable cap, $W^S\left[\mathbf{p}_0 \right]$, as well as its first forward and backward iterate, $\mathbf{p}_1$ and $\mathbf{p}_{-1}$, are also shown.  The space bounded by the stable and unstable caps is the interior of the vortex.  \textbf{c}, A scattering plot depicting the number of iterates of the map that advected particles, initially on the $z = -2$ plane below the vortex, require to cross over the $z = 2$ plane above the vortex.  This scattering plot is related to the more easily parsed escape-time plots of Fig.~\ref{ETPs}.}
		\label{HillsStreamlines}
	\end{figure*}

Our results follow from developing a deep topological understanding of special 2D surfaces within the fluid; the stable and unstable manifolds\cite{Wiggins92} attached to stagnation points.  These manifolds intersect an infinite number of times in a beautiful fractal pattern called a heteroclinic tangle (Fig.~\ref{HillsStreamlines}\textbf{b}).  The topology of the tangle has profound implications for the system dynamics.  Our technique shows how to extract this topological information and turn it into a symbolic representation of the dynamics.  This representation captures the core mixing mechanism of the fluid, generates a lower bound on the (topological) entropy, and reveals distinct mechanisms for 1D and 2D material stretching.  This work builds on the topological understanding of heteroclinic tangles in 2D\cite{Easton86,Easton98,RomKedar94,Ruckerl94a,Jung05,Collins99,Collins02a,Collins04}, namely the homotopic lobe dynamics (HLD) technique\cite{Mitchell03b,Mitchell06,Mitchell09,Mitchell12a,Sattari16,Novick12,Byrd14}, and a recent study\cite{3Dhld} showing how HLD extends to 3D for certain Òtailor-madeÓ topological constructions, defined absent any explicit dynamics.

Successfully extending the 2D HLD technique to 3D paves the way for a structural characterization of the chaotic dynamics of other volume-preserving systems, like charged particles following magnetic field lines, shifting granular media, as well as many other 3D fluid flows.  Furthermore, the HLD technique is algorithmic and amenable to automation.  This would enable the analysis of much longer timescales, and also permit analyses of bifurcating mixing mechanisms.  Finally, this current work is a spring-board to further extending the HLD technique to 4D symplectic maps derived from three-degree-of-freedom Hamiltonian systems, opening up many more applications, e.g. to chaotic atomic and molecular scattering. \\

	
\textbf{Chaotic Spherical Vortex}\\
We analyze a numerically defined, physically representative, dynamical system that exists ``in the wild'' - passive advection in a modified Hill's spherical vortex.  Hill's vortex\cite{Hill, Moffatt} (see online methods) is a well known solution to Euler's equations for an inviscid incompressible fluid (Fig.~\ref{HillsStreamlines}\textbf{a}).  Two unstable stagnation points (fixed points), $\mathbf{z}_u$ and $\mathbf{z}_{\ell}$, are connected by a 2D spherical separatrix, separating the vortex interior from its exterior.  The separatrix prevents any mixing between fluid inside and outside the vortex.
	
To induce mixing and chaos, we modify Hill's vortex by a sequence of time-periodic adjustments to the flow (see online methods).  The resulting volume-preserving advection map, $M$, which evolves an initial point $(x_i,y_i,z_i)$ forward for one period under the fluid flow to the final point $(x_f, y_f, z_f)$, becomes our primary object of study.  This advection map breaks rotational symmetry, but preserves both fixed points and reversibility.\cite{strogatz2014nonlinear} (Reversing time is equivalent to reflecting about the $xy$-plane.)  The separatrix splits into two distinct surfaces, the 2D unstable manifold $W^U$ of $\mathbf{z}_\ell$ and the 2D stable manifold $W^S$ of $\mathbf{z}_u$ (Fig.~\ref{HillsStreamlines}\textbf{b}).   (The stable/unstable manifold consists of all points that converge upon $\mathbf{z}_u$/$\mathbf{z}_\ell$ in forward/backward time.)  These two manifolds first intersect at the primary intersection curve, $\mathbf{p}_0$, defining the unstable cap $W^U\left[\mathbf{p}_0 \right]$, the piece of the unstable manifold between $\mathbf{z}_{\ell}$ and $\mathbf{p}_0$, and stable cap $W^S\left[\mathbf{p}_0 \right]$.  The vortex interior is defined as the region between the stable and unstable caps.   The stable and unstable manifolds are \emph{invariant}: each point on the manifold maps forward to another point on the manifold.  Thus, an intersection curve, e.g. $\mathbf{p}_0$, maps forward and backward to other intersection curves, e.g. $\mathbf{p}_1$ and $\mathbf{p}_{-1}$ (Fig.~\ref{HillsStreamlines}\textbf{b}). \\

	\begin{figure*}[htbp]
		\includegraphics[width = \textwidth]{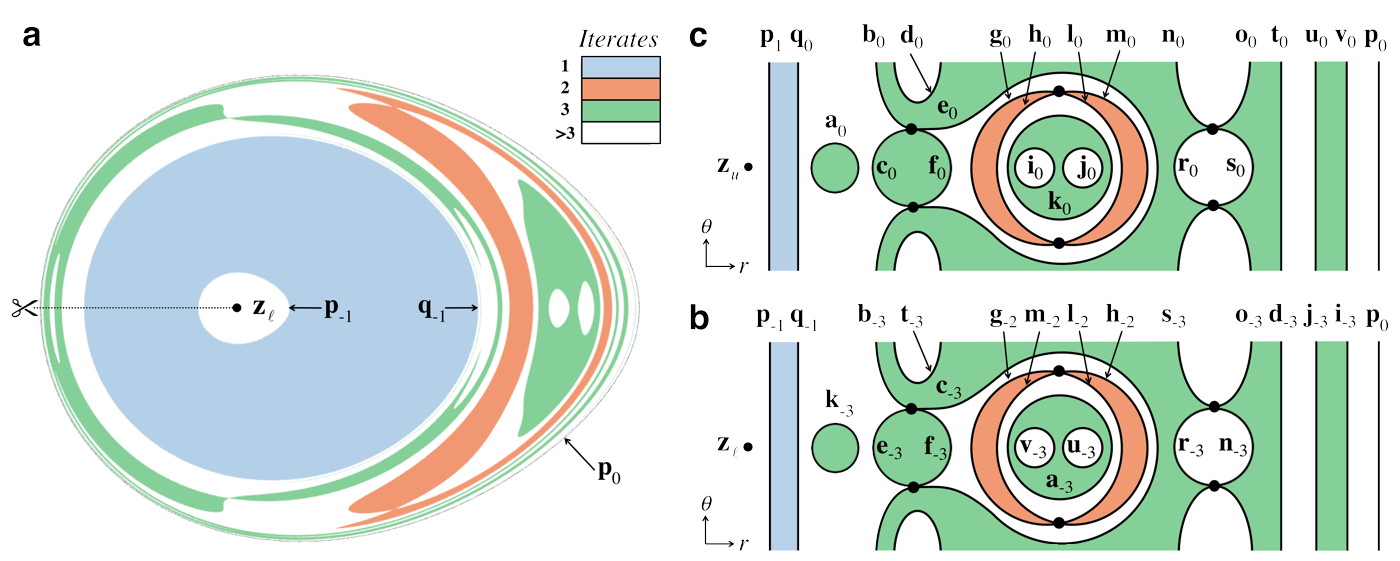}
		\caption{\textbf{Escape-Time Plots (ETPs) up to iterate three.}  \textbf{a}, Numerically computed forward ETP.  \textbf{b}, Simplified, though topologically equivalent forward ETP, with labeled intersection curves and dots indicating tangencies between $W^U$ and $W^S$.  This image can be constructed by cutting along the dotted line in \textbf{a} and folding all the topological information up, like a fan, into the rectangular region.  Escape domains that end at the top of the rectangle are to be thought of as continuing counter-clockwise around $z_{\ell}$ and connecting up with the aligned escape domain terminating on the bottom of the rectangle.  \textbf{c}, Similar cartoon of the backward ETP.  Reversibility guarantees that the forward and backward ETPs have the same pattern.}
		\label{ETPs}
	\end{figure*}

\textbf{Escape-Time Plots and Reconstructing the Tangle}\\
Due to the broken separatrix, some advected particles that originate outside the vortex will enter the vortex and subsequently escape.  We treat this as a scattering problem.  Fig.~\ref{HillsStreamlines}\textbf{c} shows the number of periods needed to pass from the plane $z = -2$ below the vortex through the plane $z = +2$ above the vortex.  Note the intricate fractal structure as a function of the two impact parameters $x$ and $y$.  For our symbolic analysis, we shall use a scattering function that is easier to interpret.  In Fig.~\ref{ETPs}\textbf{a}, the two impact parameters identify an initial point on the unstable cap, restricted to the \emph{fundamental unstable annulus} $W^U\left[\mathbf{p}_{-1},\mathbf{p}_0 \right]$ between $\mathbf{p}_{-1}$ and $\mathbf{p}_0$.  Escape is defined by leaving the vortex.  (See online methods.)  This defines the \emph{forward} escape-time plot (ETP); the \emph{backward} ETP is defined analogously for initial points in the \emph{fundamental stable annulus} $W^S\left[\mathbf{p}_{0},\mathbf{p}_{1} \right]$ moving backward in time.  Since the numerical ETP (Fig.~\ref{ETPs}\textbf{a}) includes difficult-to-see thin features, we provide clearer cartoons of the forward ETP (Fig.~\ref{ETPs}\textbf{b}) and backward ETP (Fig.~\ref{ETPs}\textbf{c}).  We shall derive the symbolic dynamics from information in these ETPs up to iterate three.
	
\begin{figure*}[htbp]
		\includegraphics[width = \textwidth]{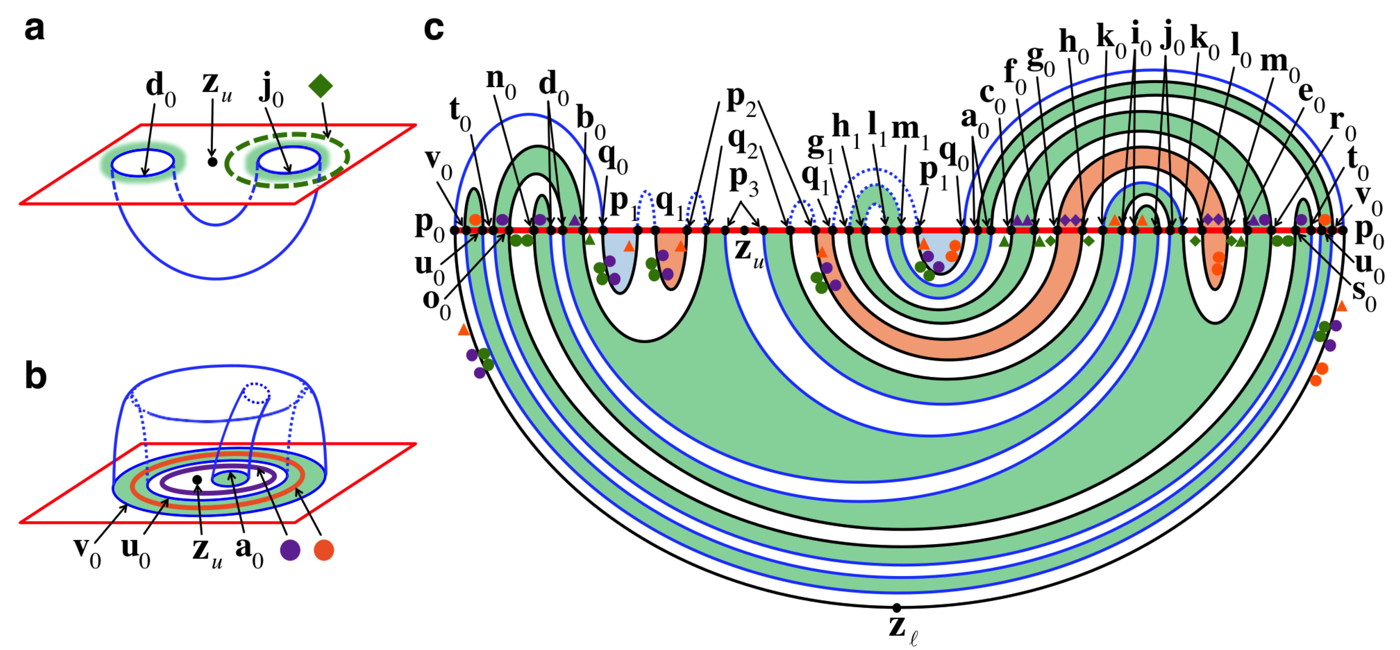}
		\caption{\textbf{Bridge Examples and Cross-Section}.  Two example bridges and some of the obstruction rings holding them up are shown: $W^U\left[\mathbf{d}_{0}, \mathbf{j}_{0} \right]$ in \textbf{a}, and $W^U\left[\mathbf{a}_{0}, \mathbf{u}_{0}, \mathbf{v}_{0} \right]$ in \textbf{b}.  The stable manifold is red and the unstable manifold is blue.  \textbf{c}, A cartoon of the tangle sliced along the plane of Fig.~\ref{HillsStreamlines}\textbf{b}, up to the third iterate of the unstable manifold.  The blue and black lines represent bridges of the unstable manifold $W^U$ and the red line represents the stable manifold $W^S$.  The black dots are the intersections between $W^U$ and $W^S$.  The colored shapes represent the obstruction rings, which enforce the minimal topology of the heteroclinic tangle and classify bridges by homotopy equivalence relative to these obstructions.  Some of the backward iterates of rings are not shown in order to reduce clutter; those omitted simply provide redundant information.  The black bridges are those with obstruction rings attached.  The volume bounded by $W^U\left[\mathbf{q}_0, \mathbf{p}_1\right]$ and $W^S\left[\mathbf{q}_0, \mathbf{p}_1\right]$ - is shaded blue.  The first and second iterates of this volume are shaded orange and green.  These help visually delineate the way different bridges fit together.}
		\label{CrossSection}
	\end{figure*}

The unstable manifold is partitioned into pieces, called \emph{bridges}, by cutting along the stable cap.  Each bridge is identified by a labeled set of intersection curves on the stable cap, e.g. $W^U\left[\mathbf{d}_{0}, \mathbf{j}_{0} \right]$ (Fig.~\ref{CrossSection}\textbf{a}) or $W^U\left[\mathbf{a}_{0}, \mathbf{u}_{0}, \mathbf{v}_{0} \right]$ (Fig.~\ref{CrossSection}\textbf{b}).  The forward ETP contains \emph{escape domains}, i.e. connected domains of constant escape time, which can be identified by their boundary labels, e.g. $W^U\left[\mathbf{a}_{-3}, \mathbf{u}_{-3}, \mathbf{v}_{-3} \right]$.  Each escape domain eventually maps to a bridge outside the vortex, e.g. $W^U\left[\mathbf{a}_{-3}, \mathbf{u}_{-3}, \mathbf{v}_{-3} \right]$ maps to $W^U\left[\mathbf{a}_{0}, \mathbf{u}_{0}, \mathbf{v}_{0} \right]$.  Similarly, each gap between escape domains, up to a given iterate, maps to an interior bridge, e.g. $W^U\left[\mathbf{d}_{-3}, \mathbf{j}_{-3} \right]$ maps to $W^U\left[\mathbf{d}_{0}, \mathbf{j}_{0} \right]$.  Thus, the boundaries of the escape domains are intersection curves between the stable and unstable manifolds.  Furthermore, each boundary curve in Fig.~\ref{ETPs}\textbf{b} is paired with a curve in Fig.~\ref{ETPs}\textbf{c} to which it maps, e.g. $\mathbf{v}_{-3}$ maps to $\mathbf{v}_0$ after three iterates.
	
To reconstruct a bridge from its boundaries, connect up each of its boundary curves with a 2D surface of genus zero (i.e. no ``handles") that doesn't intersect the rest of the unstable manifold. (Stable/unstable manifolds cannot self-intersect\cite{Wiggins92}.)  For example, the outer bridge $W^U\left[\mathbf{a}_{0}, \mathbf{u}_{0}, \mathbf{v}_{0} \right]$ (Fig.~\ref{CrossSection}\textbf{b}) looks like the top of a bisected torus, where $\mathbf{u}_{0}$ and $\mathbf{v}_{0}$ form the concentric boundary circles, connected to a tube whose other boundary is $\mathbf{a}_{0}$.  Similarly, the inner bridge $W^U\left[\mathbf{d}_{0}, \mathbf{j}_{0} \right]$ (Fig.~\ref{CrossSection}\textbf{a}) is a simple tube connecting $\mathbf{d}_{0}$ to $\mathbf{j}_{0}$.  The bridges fit inside one another like a convoluted set of nesting matryoshka dolls.  To visualize this, Fig.~\ref{CrossSection}\textbf{c} shows a cross-section of the tangle up to iterate three of the fundamental unstable annulus.  While we develop our analysis of the perturbed Hill's vortex using information up to iterate three, we have also completed this analysis up to iterate four, and include these results where applicable. \\

\textbf{Enforcing Topology: Obstruction Rings and Bridge Classes}\\
Now comes a crucial trick: we judiciously place \emph{obstruction rings} within the 3D phase space.  Bridges are now viewed as rubber sheets that can be arbitrarily distorted so long as they do not pass through the rings.  A minimal set of rings is needed to enforce the tangle topology of Fig.~\ref{CrossSection}\textbf{c}, preventing bridges from pulling through the stable cap and reducing the topological complexity.  This set is constructed by placing rings near special intersections (see online methods).  These rings are seen in profile in Fig.~\ref{CrossSection}\textbf{c}.  To appreciate the rings' function, consider the outer bridge $W^U\left[\mathbf{a}_{0}, \mathbf{u}_{0}, \mathbf{v}_{0} \right]$ (Fig.~\ref{CrossSection}\textbf{b}).  The orange ring keeps the upper half torus connecting $\mathbf{u}_{0}$ and $\mathbf{v}_{0}$ from sliding down through the stable cap, while the purple ring plays the same role for the tube connecting to $\mathbf{a}_{0}$.
	
Obstruction rings partition the set of bridges into equivalence classes.  Two bridges are of the same homotopy class, or \emph{bridge class}, if one can be continuously deformed into the other without passing through a ring.  These bridge classes form the central symbolic objects in our analysis, and their behavior under iteration constitutes a symbolic dynamical system.
	
	\begin{figure*}[htbp]
		\includegraphics[width = \textwidth]{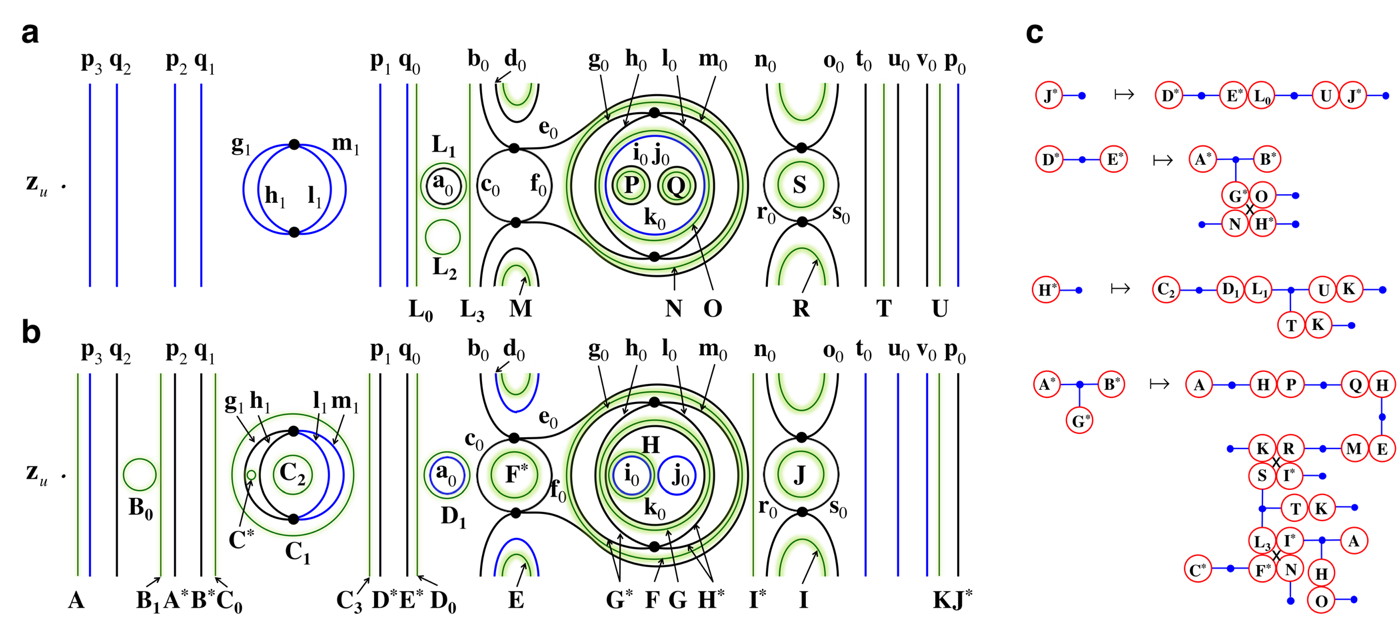}
		\caption{\textbf{Boundary Classes and Bridge Classes.} The outer, \textbf{a}, and inner, \textbf{b}, division of the stable cap, $W^S\left[\mathbf{p}_0 \right]$.  The black lines act as obstructions to deforming different boundary curves into one-another.  The inner and outer boundary classes that result are labeled with capital letters.  These are typically shown as green curves, however some of the non-recurrent boundary classes coincide with heteroclinic intersections and are shown in black.  Since every bridge class can be uniquely defined by its set of boundary classes, a convenient description of bridge classes is the ``barbell" notation of \textbf{c}.  Shown in \textbf{c} is the unstable cap $\left[\left[J^{*} \right]\right]$ and all bridges in its first three iterates.  The ``x" symbol denotes a heteroclinic tangency.}
		\label{BoundaryC}
	\end{figure*} 

Just as bridges can be specified by their boundary curves, bridge classes can be specified by the homotopy classes, or \emph{boundary classes}, of these curves.  These boundary classes are defined as follows.  The bridges with obstruction rings attached (black curves in Fig.~\ref{CrossSection}\textbf{c}) divide the stable cap into distinct regions.  The outer bridges define one division and the inner bridges another, as shown by the black curves in Figs.~\ref{BoundaryC}\textbf{a} and \ref{BoundaryC}\textbf{b}.  Two boundary curves are in the same boundary class if one can be deformed into the other without passing through the black curves.  The green curves in Fig.~\ref{BoundaryC} represent the inner and outer boundary classes needed to label the bridge classes.  For example, the bridge class of the outer bridge $W^U\left[\mathbf{a}_{0}, \mathbf{u}_{0}, \mathbf{v}_{0} \right]$ is specified by the boundary classes $\left[\left[L_1,T,U \right]\right]$, while the bridge class of $W^U\left[\mathbf{j}_{0}, \mathbf{d}_{0} \right]$ is specified by $\left[\left[H, E \right]\right]$.  The ``barbell" pictograms of Fig.~\ref{BoundaryC}\textbf{c} are useful graphical representations of bridge classes.  The labeled red circles represent the boundary classes on the stable cap, and the blue lines connecting them represent the bridge class itself.  \\
		
	\begin{figure*}[htbp]
		\includegraphics[width=\linewidth]{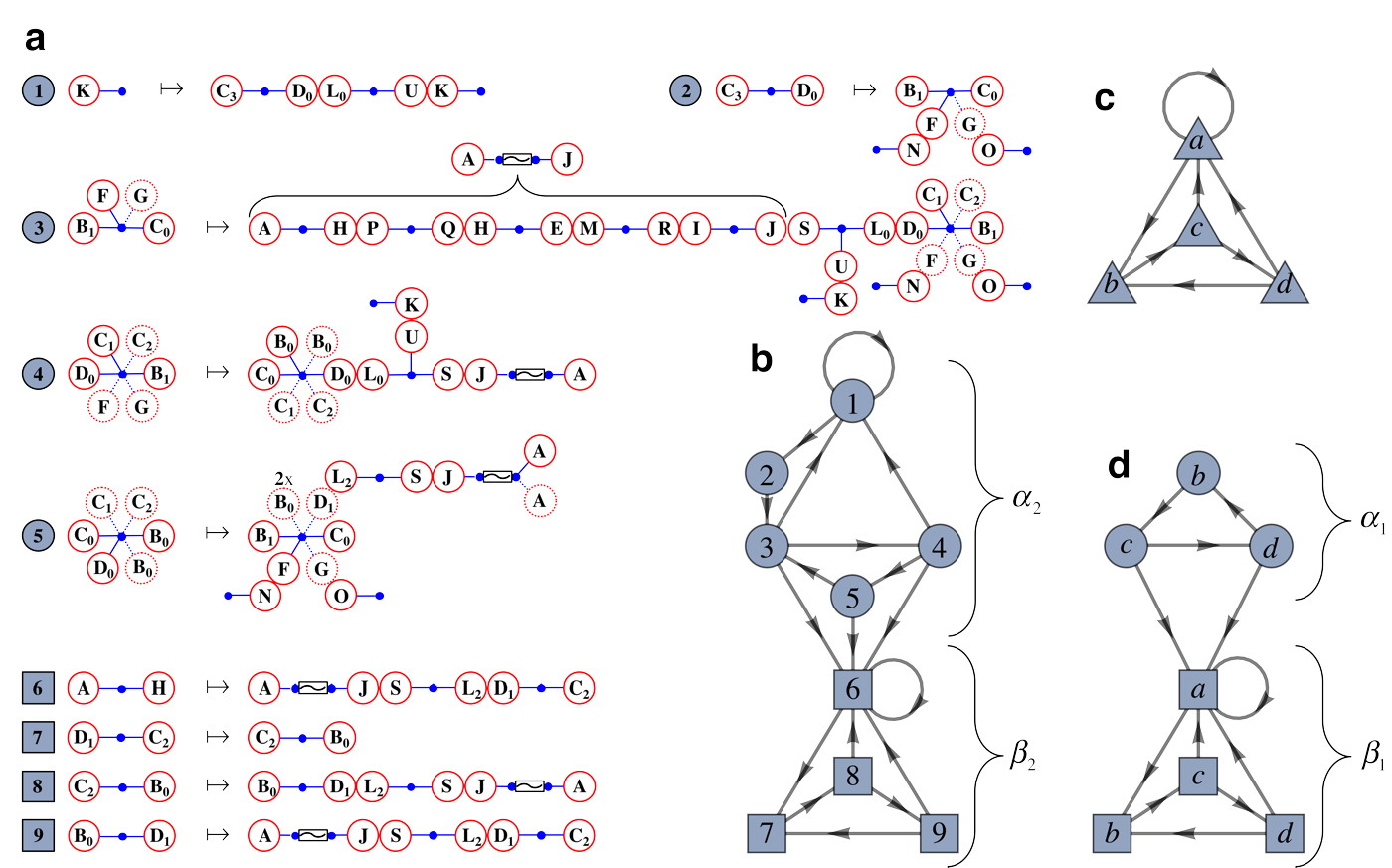}
		\caption{\textbf{Recurrent Bridge Classes and Symbolic Dynamics.}  \textbf{a}, The active recurrent bridge classes and their iterates.  Boundary classes marked with a dotted circle represent inert boundary classes (see online methods).  Bridge classes such as $\left[\left[B_1,F,G,C_0\right]\right]$ and $\left[\left[B_1,F,G,C_0,D_1,B_0,B_0\right]\right]$ are considered to be equivalent, as they differ by inert boundary classes only.  \textbf{b}, Transition graph representing the symbolic dynamics of 2D bridge classes.  Note the two strongly connected components (SCCs): vertices with circles correspond to the 2D stretching mechanism, $\alpha_2$, and those with squares correspond to the 1D stretching mechanism, $\beta_2$.  \textbf{c}, Transition graph for the 1D bridge classes.  \textbf{d}, The 1D bridge class dynamics with the added distinction of whether the 1D bridge is embedded in a 2D bridge that is part of the 2D stretching mechanism (circles - $\alpha_1$), or part of the 1D stretching mechanism (squares - $\beta_1$).}
		\label{BigSymbolicGraph}
	\end{figure*}

\textbf{Dynamics of the Bridge Classes}\\
Evolving forward one period, each bridge maps to a set of alternating inner and outer bridges, glued together at their common boundary curves.  Crucially, this idea extends to bridge classes (see Fig.~\ref{BoundaryC}\textbf{c}).  The left side shows the class, in barbell form, of every bridge in the tangle up to iterate two.  The right side shows what each of these classes maps to, depicted as a gluing together of bridge classes.  The right side of Fig.~\ref{BoundaryC}\textbf{c} contains new classes not present on the left.  We must compute how these new classes map forward (see online methods).  Repeating this procedure, we eventually obtain a set of bridge classes that is closed under iteration.  Computing how bridge classes map forward is much of the ``real work" of the HLD technique.
	
Some classes, e.g. $\left[\left[ D^*,E^*\right]\right]$, only occur once when repeatedly iterating the unstable fundamental annulus forward.  These transient, or non-recurrent, classes convey no long-term topological information, and are ignored.  Similarly, we ignore all inert classes, those that map to exactly one class under repeated iterations.  For example, every outer bridge class is inert.

Figure~\ref{BigSymbolicGraph}\textbf{a} shows the remaining non-inert, or \emph{active}, classes that are also recurrent, and their iterates.  The graph in Fig.~\ref{BigSymbolicGraph}\textbf{b} records the allowed transitions between bridge classes: each vertex represents the correspondingly numbered class in Fig.~\ref{BigSymbolicGraph}\textbf{a}, and the directed edges show which classes are produced upon one iterate.  The graph can also be presented as a transition matrix, $\mathbf{A}$, where $A_{ij}$ is the number of directed edges connecting vertex $j$ to vertex $i$.	
	
The HLD analysis can be repeated for 1D bridges, paths beginning and ending on the stable cap and embedded in a 2D bridge.  These 1D bridges can similarly be grouped into 1D bridge classes based on how they wrap around the obstruction rings.  The forward iterates of 1D bridge classes can then be obtained by iterating forward the 2D bridge classes in which they are embedded.  Fig.~\ref{BigSymbolicGraph}\textbf{c} shows the transition graph of the resulting symbolic dynamics.  \\

\textbf{Topological Entropy and Stretching Rates}\\
The complexity of mixing in the vortex can be characterized by its topological entropy, which measures the exponential growth rate in the number of ``distinguishable" orbits as a function of time \cite{Bowen,Young03}.  Intimately related to this is the topological entropy of the symbolic dynamics, or symbolic entropy, which is computed as the log of the largest eigenvalue of the transition matrix.  Since the symbolic dynamics represent the minimal topology forced by our knowledge of the tangle, its symbolic entropy is a strict lower bound to the full topological entropy of the advection map.  Importantly, we can systematically increase this lower bound by including information about the tangle at successively higher iterates.  For instance, our knowledge of the tangle up to iterate three (Fig.~\ref{BigSymbolicGraph}\textbf{b}) gives a symbolic entropy of $\ln\left(1.6956\right)$, while the extra information gleaned at iterate four increases this to $\ln\left(2.1106\right)$ (see online methods).  With increasing information about the tangle, the symbolic entropy will converge from below on the full topological entropy.  We have directly computed this full topological entropy to lie between $\ln\left(2.7114\right)$ and $\ln\left(2.8210\right)$ using an independent approach \cite{HuntOtt}, thus confirming our symbolic calculation as a strict lower bound.
	
Topological entropy also measures the exponential stretching rate of the area or length of material advected in the fluid.  The symbolic entropies of the graphs in Figs.~\ref{BigSymbolicGraph}\textbf{b} and \ref{BigSymbolicGraph}\textbf{c} give lower bounds on the stretching rates of 2D material surfaces and 1D material lines, respectively.  Differing 2D and 1D stretching rates is a unique possibility of the 3D HLD analysis, as seen in ref.~\cite{3Dhld}.  However, in this example the 1D and 2D stretching rates are \emph{both} $\ln\left(1.6956\right)$ for the iterate-three analysis and $\ln\left(2.1106\right)$ for the iterate-four analysis.  Rather than being a coincidence, the equality of these two rates reflects a deeper structural similarity between the 1D and 2D stretching mechanisms, as discussed next.\\

\textbf{Two Stretching Mechanisms}\\
The transition graph (Fig.~\ref{BigSymbolicGraph}\textbf{b}) has two strongly connected components (SCCs), labeled $\alpha_2$ and $\beta_2$, i.e. subgraphs in which every vertex is reachable from every other vertex.  Each SCC represents a distinct stretching mechanisms at work in the vortex.  This distinction is revealed by a closer look at the dynamics of 1D bridge classes.  Let $\alpha_1$ ($\beta_1$) be the subset of 1D bridge classes that can be embedded in an $\alpha_2$ ($\beta_2$) 2D bridge class.  1D bridge classes in Fig.~\ref{BigSymbolicGraph}\textbf{c} common to both $\alpha_1$ and $\beta_1$ can be symbolically split to enable an ``unfolded" version of the 1D symbolic dynamics graph, Fig.~\ref{BigSymbolicGraph}\textbf{d}.  This splitting allows us to examine the stretching forced on 1D material lines by $\alpha_2$ and $\beta_2$ separately.
	
Since Fig.~\ref{BigSymbolicGraph}\textbf{c}, $\beta_1$ (Fig.~\ref{BigSymbolicGraph}\textbf{d}), and $\beta_2$ (Fig.~\ref{BigSymbolicGraph}\textbf{b}) all show identical symbolic dynamics, the stretching of the 2D $\beta_2$ bridge classes is fundamentally 1D.  Indeed, each bridge in the $\beta_2$ SCC is a simple tube, which stretches out along its axis to multiple such tubes under iteration, as can be seen in equations $6-9$ of Fig.~\ref{BigSymbolicGraph}\textbf{a}, where each barbell is mapped to a chain of barbells.  On the other hand, since the dynamics of $\alpha_1$ are trivial, i.e. have zero topological entropy, while those of $\alpha_2$ are more complex, the stretching of $\alpha_2$ bridge classes is irreducibly 2D in nature.  As an example, consider the $\left[\left[ K\right]\right]$ bridge class, labeled 1 in Fig.~\ref{BigSymbolicGraph}\textbf{b}, which is topologically an interior cap.  While it is the central symbol in $\alpha_2$ and participates in producing non-zero topological entropy, it does not force any 1D stretching, as any 1D bridge embedded in such a cap is contractible to a point.  On the other hand, 2D bridges in $\left[\left[ K\right]\right]$ are pushed down against the unstable cap and stretched out radially away from the lower fixed point.  This fundamentally 2D behavior is also seen in equations $1-5$ of Fig.~\ref{BigSymbolicGraph}\textbf{a}, which exhibit branching not possible for 1D stretching.  These two stretching mechanisms exist in the iterate-four dynamics as well. (See online methods).

While the two stretching mechanisms, $\alpha_2$ and $\beta_2$, are different in terms of the dimensionality of the structures which drive their stretching rates, their symbolic entropies are identical.  More tellingly, when taken as formal symbolic dynamical systems, i.e. bi-infinite shifts of finite type, we discovered a strong shift equivalence between $\alpha_2$ and $\beta_2$, implying that their dynamics are identical up to topological conjugacy\cite{LindMarcus, Kitchens}.  These facts holds at iterate-four as well.  Far from being a coincidence, we conjecture that this equivalence is due to an underlying duality between forward-time 1D dynamics and backward-time 2D dynamics, and will appear in the dynamics forced by the tangle up to any iterate.\\
		
	\begin{figure*}[htbp]
		\includegraphics[width = \textwidth]{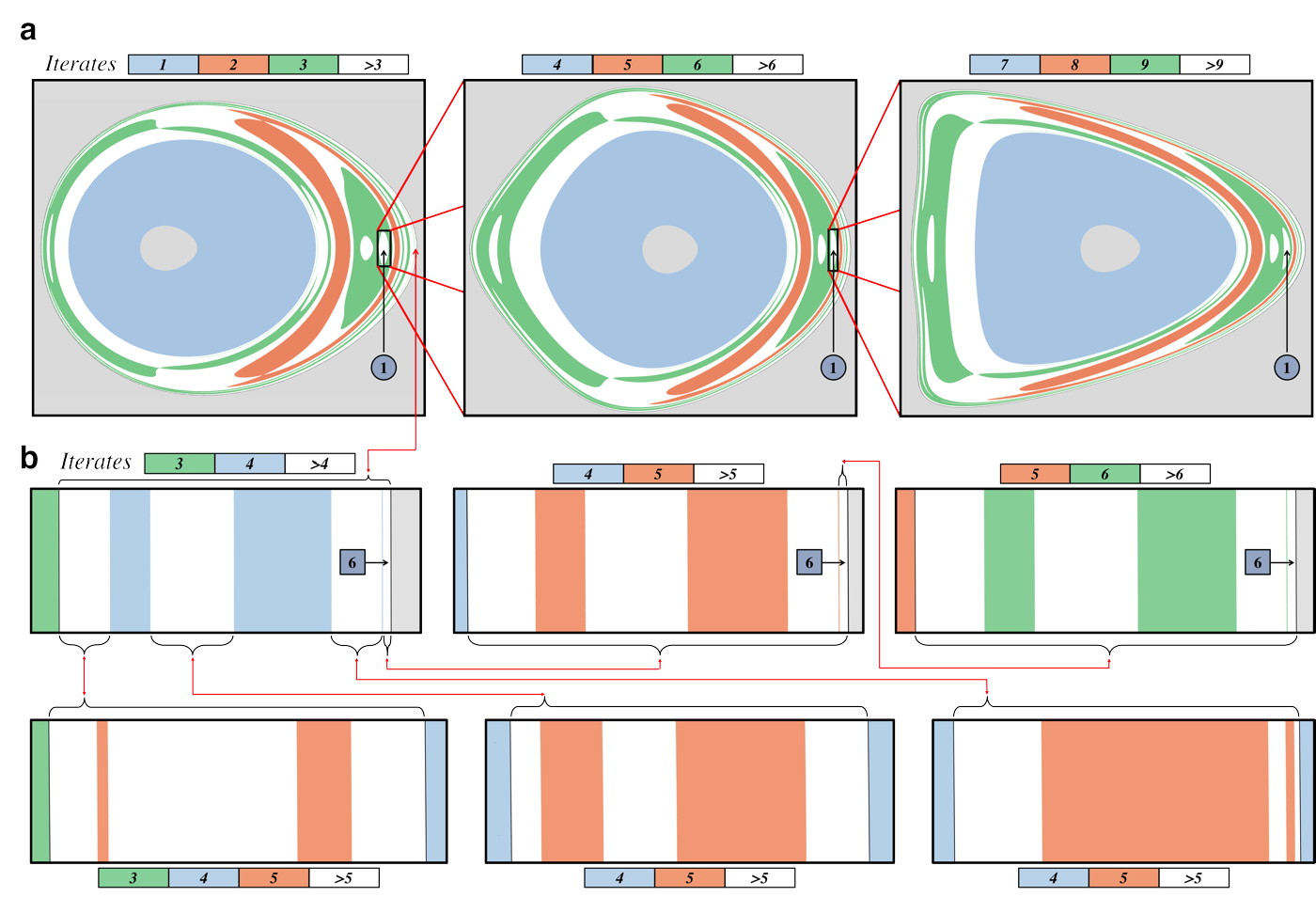}			
		  \caption{\textbf{Fractal Structure} Numerical self-similar fractal regions.  \textbf{a}, A complicated, fundamentally 2D, fractal structure, which is represented in the Fig.~\ref{BigSymbolicGraph}\textbf{b} graph by the 3-cycle $1 \rightarrow 2 \rightarrow 3$.  \textbf{b}, The three boxes on the top row each show three annular bands which repeat indefinitely upon zooming into the right-most unescaped region, described by bridge class $\left[\left[H, A \right]\right]$ or symbol 6 in Fig.~\ref{BigSymbolicGraph}.  The zooming is represented in the Fig.~\ref{BigSymbolicGraph}\textbf{b} graph by the edge connecting 6 to itself.  Note that all three bands in this 1D stretching mechanism fractal are predicted by the symbolic dynamics only after incorporating information up to iterate four.  In the bottom row, from left to right, the minimal number of iterate-five escape domains forced by iterate-four knowledge are 0, 0, and 2.  The extra observed escape domains in the first two bottom boxes illustrate that the actual structure can be more complicated than the forced structure, and that higher iterate information can be included in the analysis to enlarge the forced dynamics.}
		  \label{FractalStructure}
	\end{figure*} 	

\textbf{Fractal Structure}\\
We started with scattering information in the ETPs, from which we derived a symbolic dynamics (Figs.~\ref{BoundaryC}\textbf{c} and \ref{BigSymbolicGraph}\textbf{a}) for the minimal topological structure.  This process can be reversed, and we can use the equations Fig.~\ref{BoundaryC}\textbf{c} and Fig.~\ref{BigSymbolicGraph}\textbf{a}, to fully reconstruct the ETPs.  More importantly, we can predict new, topologically-forced, escape domains at higher iterates.  These predictions provide a direct way to validate the symbolic dynamics by comparing them to numerical ETP results at higher iterates.  Particularly interesting as test cases are cycles in the transition graph, which generate fractal self-similar patterns in the ETP.  Each symbol represents a motif in the overall fractal, and every gap in the ETP that corresponds to this symbol will contain the same pattern at higher iterates.  As an example, consider the 3-cycle  $1 \rightarrow 2 \rightarrow 3$ in Fig.~\ref{BigSymbolicGraph}\textbf{b}, or more precisely, the analogous cycle in the iterate-four symbolic dynamics (see online methods).  Every symbol ``1" generated upon traversing this cycle once corresponds to a gap in each of the three panels of Fig.~\ref{FractalStructure}\textbf{a}.  Zooming into this gap replicates this motif ad infinitum.  Since this cycle is within the $\alpha_2$ stretching mechanism, the associated motif is fundamentally 2D.  Contrast this with the 1-cycle of 6 repeated within the $\beta_2$ stretching mechanism, whose associated fractal is shown in the top three boxes of Fig.~\ref{FractalStructure}\textbf{b}.  This fractal structure is a zoomed-in portion of concentric annuli, which highlights its 1D nature.  In both example fractals, the symbolic dynamics predict the exact fractal structure in the numerical ETPs.  Note that iterate-four information was needed to produce such accurate symbolic dynamics, demonstrating the necessity ---and power--- of folding in new information at higher iterates.  Indeed, at iterate five and higher, escape domains exist that are not predicted by iterate-four knowledge, e.g. the lower row of Fig.~\ref{FractalStructure}\textbf{b} shows extra, unpredicted bands at iterate five.  Aside from being beautiful manifestations of the complexity inherent in the vortex, this fractal analysis reinforces the important idea that our finite knowledge of the scattering data has forced the existence of an infinite succession of predictable structures in the tangle. \\

\pagebreak
\textbf{Methods} \\

\small
\textbf{Hill's Spherical Vortex.}  This well known fluid flow is given by the stream-function\cite{Hill, Moffatt}	
	\begin{equation}
		\psi\left(r,\theta \right)  = \\
		\begin{cases}
			\frac{1}{2}U\left(1-\frac{a^3}{r^3}\right)r^2\sin^2\theta, &\left(r > a\right), \\
			-\frac{3}{4}U\left(1-\frac{r^2}{a^2}\right)r^2\sin^2\theta,  &\left(r < a\right),
		\end{cases}
		\label{StreamFunc}
	\end{equation}
where $\left(r,\theta\right)$ are the radial and azimuthal spherical coordinates (no equatorial dependence), $a$ is the vortex radius (set to $a = 1$), and $U$ is the magnitude of the uniform velocity field far from the vortex (set to $U = -1.2573$).  Changing to cylindrical coordinates, the velocity field is $\dot{\rho} = - \rho^{-1}\partial \psi/\partial z$ and $\dot{z} = \rho^{-1}\partial \psi/ \partial \rho$.  This velocity field,
		\begin{equation}
			\dot{\rho} =  
			\begin{cases}
				\frac{3U}{2}\frac{a^3\rho z}{\left(z^2+\rho^2\right)^{\frac{5}{2}}}  &\left(r > a\right),\\
				\frac{3U}{2}\frac{\rho z}{a^2}  &\left(r < a\right),
			\end{cases}
			\label{VelEqrho}
		\end{equation}
		\begin{equation}
			\dot{z} =  
			\begin{cases}
				\frac{U}{2}\left(1 + \frac{a^3}{\left(z^2+\rho^2\right)^{\frac{3}{2}}} + \frac{3}{2}\frac{a^3\rho^2}{\left(z^2+\rho^2\right)^{\frac{5}{2}}} \right)  &\left(r > a\right),\\
				-\frac{3U}{2}\left(1-2\frac{\rho^2}{a^2}-\frac{z^2}{a^2}\right)  &\left(r < a\right),
			\end{cases}
			\label{VelEqz}
		\end{equation}
is volume preserving. \\

\textbf{Modifications to the Fluid Flow.}  We first create a map, $H$, by integrating initial points $\left(x,y,z\right)$ over the time interval $\left[0,1\right]$, using the velocity field, Eq.~(\ref{VelEqrho}) and (\ref{VelEqz}).  Next we compose $H$ with a series of maps, each intended to break a particular symmetry while maintaining certain features.  To break the separatrix, we use the map, $L_z\left(x,y,z\right) = \left(x,y,z+\epsilon\left(x^2+y^2\right) \right)$, where the strength of the perturbation is $\epsilon = 0.75$.  To break the rotational symmetry about the z-axis, we apply a rotation about the y-axis, $R_y\left(\theta\right)$.  The rotation angle, $\theta$, is position dependent, $\theta\left(r\right) = 2\pi\delta_y\left(a-r\right)/\left(1+r^2\right)$, where $\delta_y = 0.3$.  Finally, we introduce an additional rotation about the z-axis, $R_z\left(\theta\right)$, with $\theta\left(r\right) = 2\pi\delta_z\left(a-r\right)/\left(1+r^2\right)$, where $\delta_z = 0.2$.  To ensure the final map is reversible, i.e. $M^{-1} = S \circ M \circ S$, where $S\left(x,y,z\right) = \left(x,y,-z\right)$, we compose the functions as
\begin{equation*}
M \equiv R_z^{-1} \circ R_y \circ L_z \circ H \circ L_z \circ R_y \circ R_z.
\end{equation*} 
This map is volume-preserving and also preserves the fixed points $z_u = \left(0,0,1\right)$ and $z_\ell = \left(0,0,-1\right)$. \\

	\begin{figure*}[htbp]
		\includegraphics[width = \textwidth]{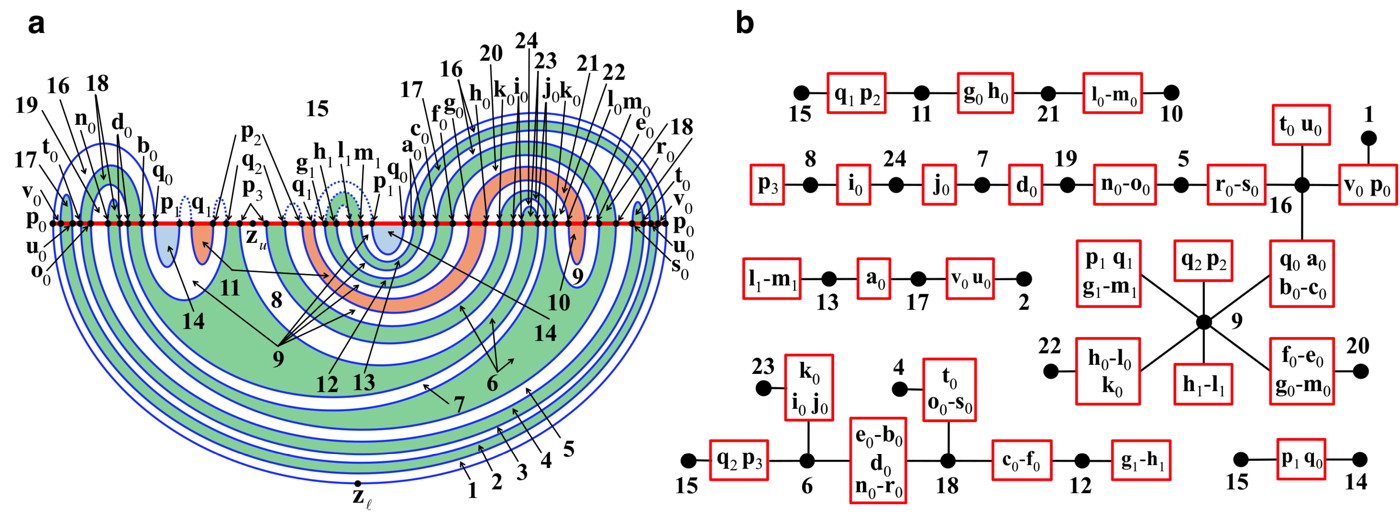}
		\caption{\textbf{Secondary Division and Connection Graphs}. \textbf{a}, A cross-section of the tangle up to third iterate where each bridge (blue lines) is part of the secondary division.  The numbers label the connected regions created by cutting phase space along these bridges.  These regions can share common boundaries on the stable fundamental domain.  \textbf{b}, The connection graph formed by connecting the numbered regions (dots) with their neighbors across the fundamental stable annulus.  The red boxes denote the 2D domains forming a common boundary between the regions.  These domains are labeled by the intersection curves forming their boundary (Fig.~\ref{ETPs}\textbf{c}).  The forward iterate of each bridge class is forced to lie in one of the connected components of this graph.}  
		  \label{ConnectionGraphs}
	\end{figure*}

\textbf{Numerical Calculations.}  To numerically generate the stable and unstable caps, we start with linear approximations (small disks) of the stable and unstable manifolds near the two fixed points.  These disks, represented by a high-density random sampling of points, are iterated (forward for $W^U$ and backward for $W^S$), until they intersect.  These caps define the vortex boundary, and thereby enable us to build the escape-time plot (ETP).  

The forward ETP begins as a small annulus on $W^U$ about the lower fixed point, $\mathbf{z}_{\ell}$, with a difference between its outer and inner radii large enough to capture a whole pre-iterate of the fundamental unstable annulus.  Each of the randomly sampled points on this annulus are iterated forward until they have either exited the vortex, or reached a predetermined maximum number of iterates.  We associate with each initial point the number of iterates it took to escape, i.e. its escape time.   Next, we compute the Delaunay triangulation of these initial points.  Escape domains are now identified as sets of points sharing a common escape time and all mutually connected through triangulation paths that contain only points in this set.  The gaps are analogously identified for contiguous regions of points that have yet to escape at the maximum iterate.  It should be noted that since escape domains are identified within gaps of smaller maximum iterate, gaps and escape domains form a natural tree structure which is convenient for data organization.

Domain boundaries are identified as triangles whose vertices do not share a common escape time.  These triangles, and select adjacent triangles, can then be infilled with new points (which are themselves iterated forward) and re-triangulated to enable an iterative refinement of the boundaries.  This allows us to increase the resolution of the forward ETP.  Due to reversibility, the gaps and escape domains of the backward ETP are geometrically identical to those of the forward ETP.  To create a complete backward ETP we must label each intersection curve as the forward iterate of a specific intersection curve in the forward ETP.  To do this, we iterate every domain boundary in the forward/backward ETP forward/backward an appropriate number of times, and note which pairs coincide in phase space.  Where this process was not possible, we used topological self-consistency to make the identification. \\

\textbf{Heteroclinic Tangencies and Time-Reversal Symmetry.}  In addition to the transverse intersections between the stable and unstable manifolds that form closed curves, the dynamics also exhibit a countably infinite number of heteroclinic tangencies (the dots in the ETPs of Figs.~\ref{ETPs}\textbf{b} and \ref{ETPs}\textbf{c}).  At each of these sites, the two manifolds locally intersect like a flat plane intersecting a saddle, forming an ``x" pattern.  Thus, the intersection curves in each ETP are either simple closed curves, or collections of curve segments whose ends terminate at a tangency.  Interestingly, the existence of reversibility ensures the robustness of these tangencies.  Any perturbation to the unstable or stable manifold local to a tangency that removes this tangency necessarily violates reversibility.  Thus, tangencies are robust under small changes to the parameter set $\left(A,\epsilon, \delta_y, \delta_z\right)$, and are only removed through more global topological changes to the tangle.	\\

	\begin{figure*}[htbp]
		\includegraphics[width = \textwidth]{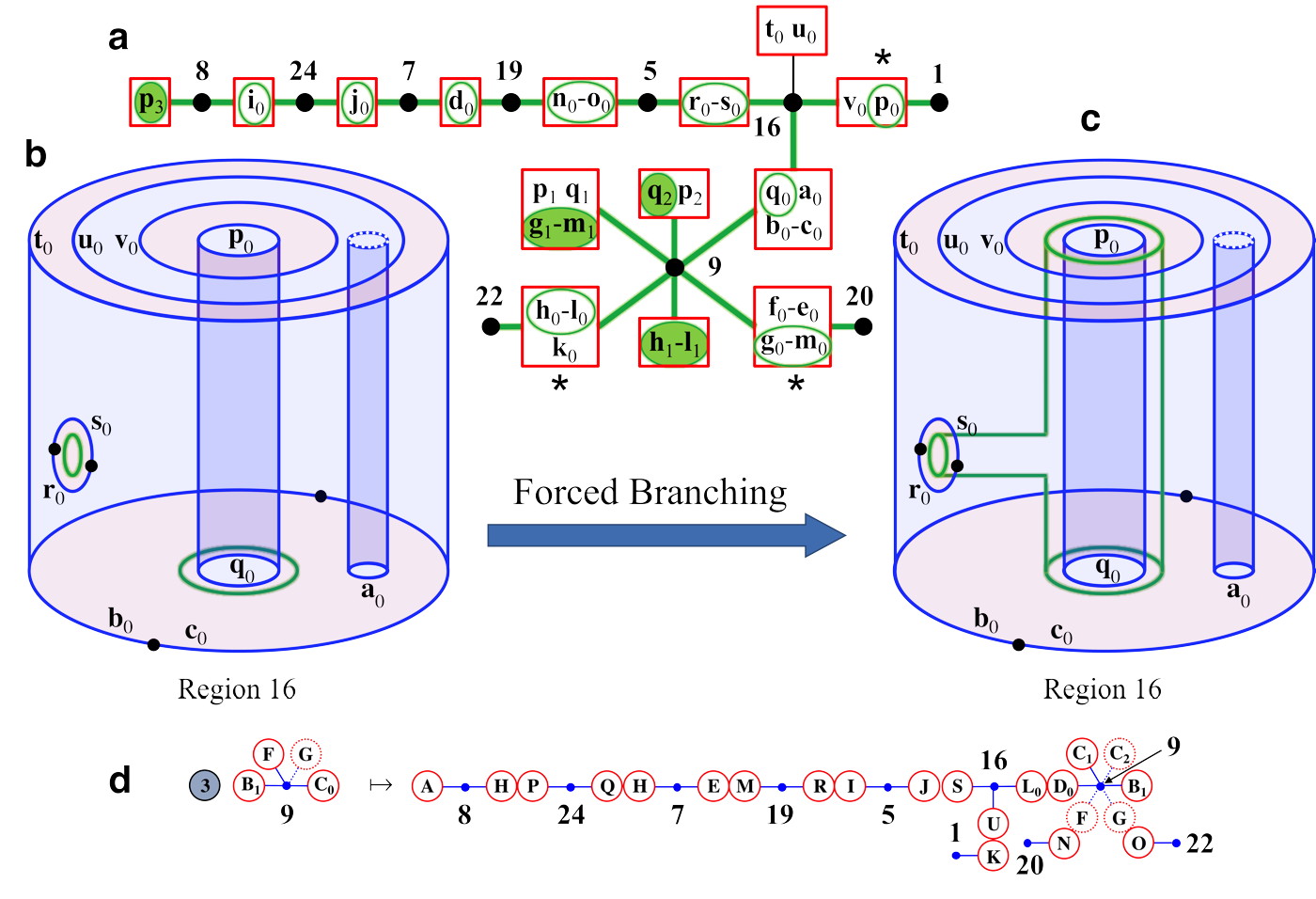}
		\caption{\textbf{Forcing Example}.  \textbf{a}, One of the connected components of the connection graph, Fig.~\ref{ConnectionGraphs}\textbf{b}.  The green lines indicate the path that the forward iterate of the bridge class $[B_1,C_0,F,G]$ is forced to traverse through the associated phase space regions.  The green circles represent the intersection of this piece of the unstable manifold with the stable cap, where each intersection curve can be seen as a small perturbation away from the existent intersection curves represented by the encircled symbols.  The intersection curves shaded in solid green are the forward iterates of the initial boundary classes, $B_1,C_0,F,G$, while those marked with an asterisk are intersection curves newly forced by the specific topology of the adjacent region.  \textbf{b}, A topological representation of region 16.  Blue surfaces are part of the unstable manifold, red surfaces are part of the stable cap, and the labeled blue lines are the intersection curves.  For reference, the bridge $[\mathbf{a}_0,\mathbf{u}_0,\mathbf{v}_0]$ is also depicted in Fig.~\ref{CrossSection}\textbf{b}.  The two green ellipses are boundary curves that are forced to exist as part of the minimal path in the connection graph which connects the four shaded green boundary curves.  However, as can be seen in \textbf{c}, these two initial boundary curves cannot be connected in region 16 without forcing a branching in the unstable manifold, which necessitates an additional boundary curve about $\mathbf{p}_0$.  \textbf{d}, The bridge class iterate equation, from Fig.~\ref{BigSymbolicGraph}\textbf{a}, that corresponds to this example.  The green path in \textbf{a} yields the concatenation of bridge classes on the right-hand side of \textbf{d}, where the numbers label the region in which each bridge class lives.  Note that the bridge class $[K]$ (symbol 1 in Fig.~\ref{BigSymbolicGraph}\textbf{a}) forced from the above analysis of region 16, is a central symbol in the symbolic dynamics.  Removing this symbol eliminates all the entropy from the $\alpha_2$ stretching mechanism.}  
		  \label{Forcing}
	\end{figure*} 

	\begin{figure*}[htbp]
		\includegraphics[width = \textwidth]{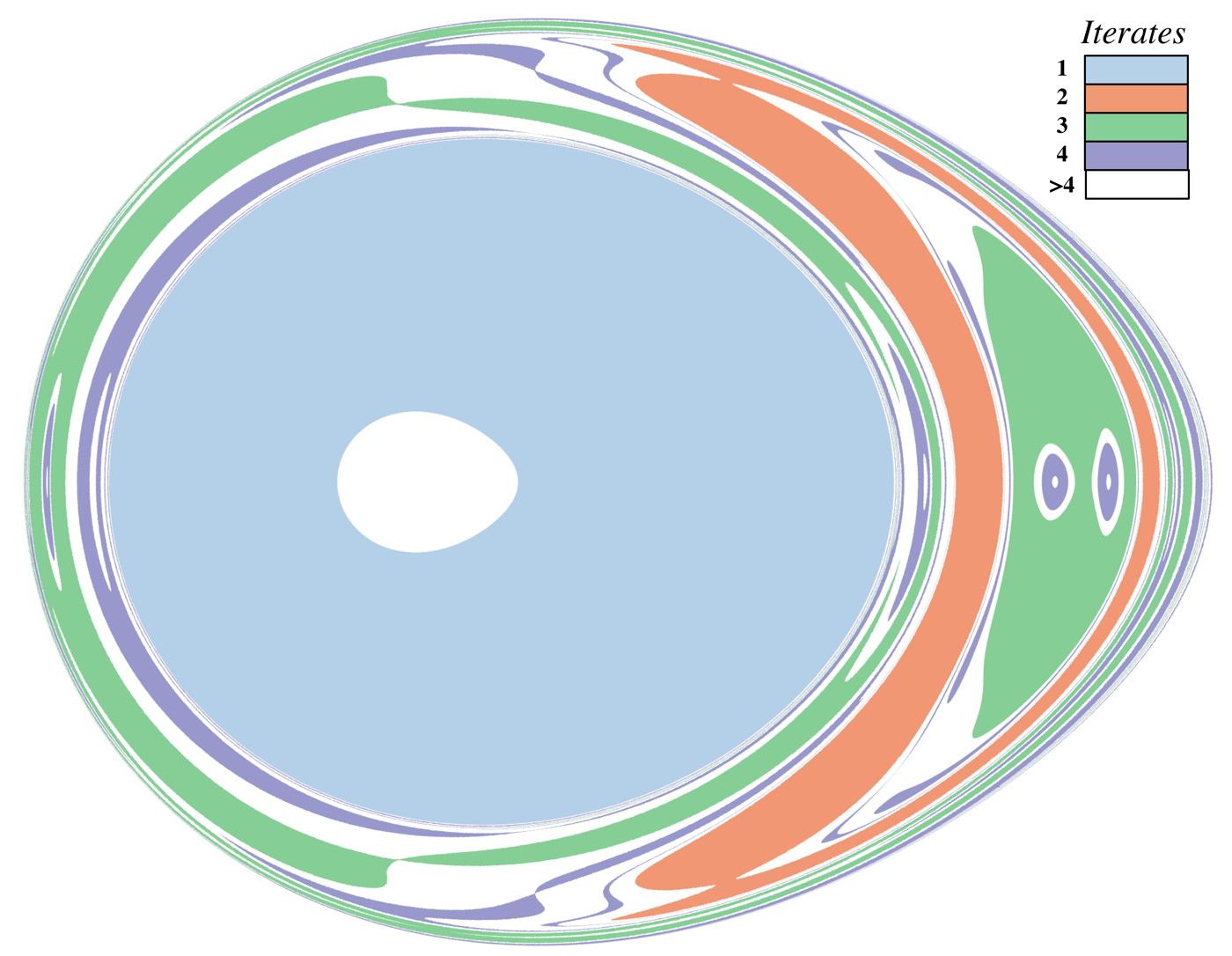}
		\caption{\textbf{Numerical Escape-Time Plot up to Iterate Four}.  This is similar to Fig.~\ref{ETPs}\textbf{a}, with the addition of fourth iterate information.}  
		\label{ETP4num}
	\end{figure*} 

	\begin{figure*}[htbp]
		\includegraphics[width = \textwidth]{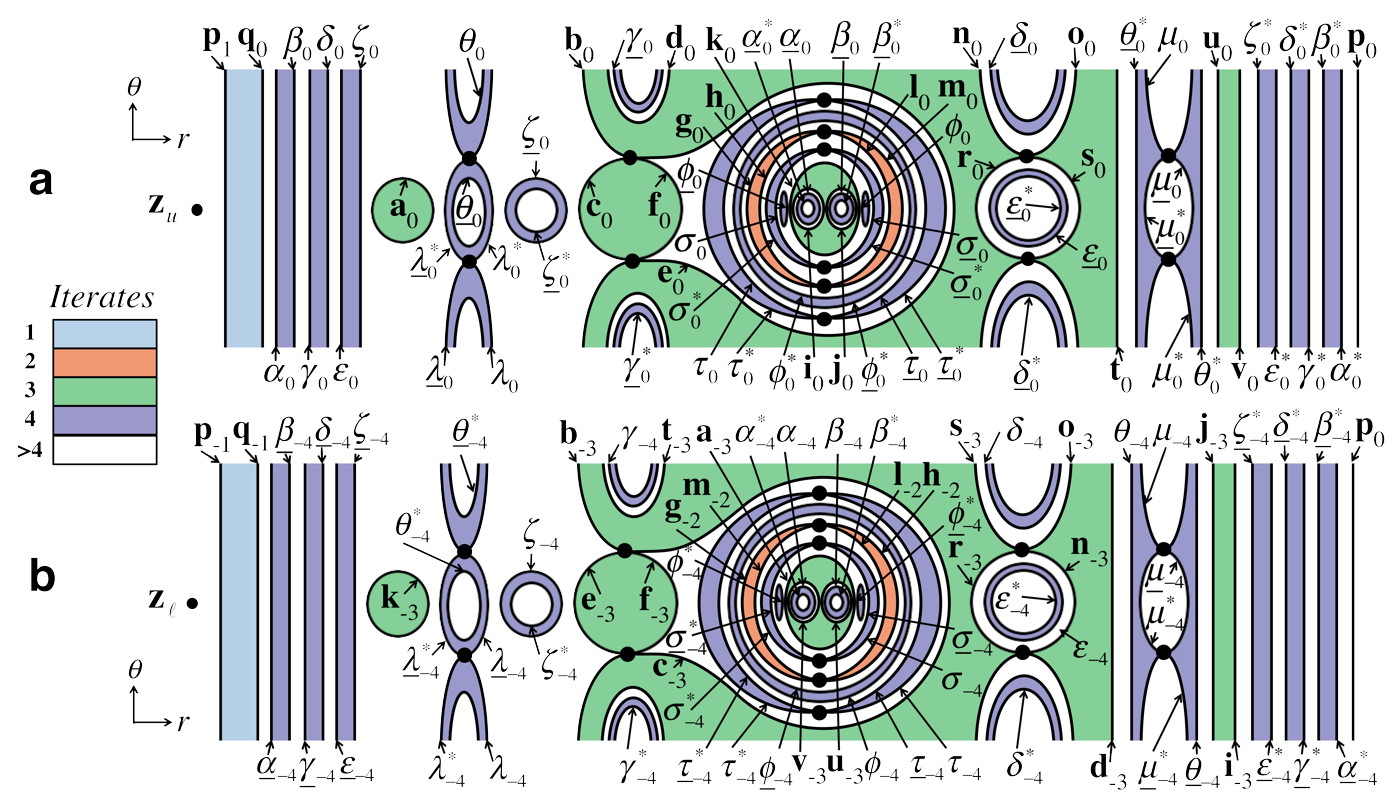}
		\caption{\textbf{Simplified, Though Topologically Equivalent, Escape-Time Plots up to Iterate Four}.  \textbf{a}, The backward ETP.  \textbf{b}, The forward ETP.  In both cases, the intersection curves at fourth iterate are labeled with greek letters.  The underlined, starred, underlined and starred, and regular variations of a letter are distinct intersection curves.}  
		\label{ETP4}
	\end{figure*}

\textbf{Obstruction Rings and Pseudoneighbors.}  We algorithmically identify the set of obstructions rings, which enforce the given tangle topology and determine the equivalence relation defining the bridge classes, by first identifying specific intersection curves called \emph{pseudoneighbors}.  Pairs of intersection curves are pseudoneighbors if their iterates in the stable annulus can be connected by a path  without crossing another intersection curve, and similarly for the unstable annulus.  Every forward and backward iterate of a pseudoneighbor pair is likewise a pseudoneighbor pair.  There is also the possibility of having a self-pseudoneighbor, a single intersection curve that bounds disks of $W^U$ and $W^S$, that contain no other intersection curves.  This definition of pseudoneighbors applies to intersection curves emanating from tangencies as well.  Pseudoneighbors are essentially the minimal set of intersection curves whose existence force the existence of every other intersection curve up to a given iterate.  The pseudoneighbors that lie in the stable annulus, seen in Fig.~\ref{ETPs}\textbf{c}, are $\left(\mathbf{u}_0,\mathbf{v}_0\right)$, $\left(\mathbf{i}_0, \mathbf{j}_0\right)$, $\left(\mathbf{g}_0-\mathbf{h}_0-\mathbf{l}_0-\mathbf{m}_0\right)$, $\left(\mathbf{b}_0-\mathbf{c}_0-\mathbf{e}_0-\mathbf{f}_0\right)$, and $\left(\mathbf{n}_0-\mathbf{o}_0-\mathbf{r}_0-\mathbf{s}_0\right)$.  The intersection curves separated by dashes indicate that they are connected through a heteroclinic tangency and should be considered as a single pseudoneighbor.  Thus, the final three are self-pseudoneighbors associated with the three pairs of heteroclinic tangencies.

The obstruction rings are placed infinitesimally close to either intersection curve of a pseudoneighbor pair, perturbed away from the curve into the quadrant defined by the pieces of $W^U$ and $W^S$ connecting the pair of pseudoneighbors.  Self-pseudoneighbors engender obstruction rings that are perturbed toward the interior of the region bounded by disks of $W^U$ and $W^S$ that have this self-pseudoneighbor as a mutual boundary.  The obstruction rings associated with self-pseudoneighbors attached to tangencies are not rings, but copies of all the intersection curves mutually connected through tangencies, perturbed above and below the pseudoneighbor.  Obstruction rings associated with pseudoneighbor pairs that lie on the stable cap outside of the stable annulus are not included, as they would serve to distinguish inert bridge classes that play no role in the active symbolic dynamics of Fig.~\ref{BigSymbolicGraph}\textbf{b}.	\\

\textbf{The Primary Division and Boundary Classes.}  Each bridge class is identified by its collection of boundary classes.  In order to define boundary classes, we define an inner and outer division of the stable cap, which act as the equivalence-class-defining obstructions.  These divisions are, in turn, restrictions of a division of the full 3D phase space, the \emph{primary division}, to the stable cap.  The primary division is constructed by dividing up phase space using the stable cap and every bridge that has an obstruction ring adjacent to it.  The black lines in Fig. \ref{CrossSection}\textbf{c} show the primary division, while Figs. \ref{BoundaryC}\textbf{a} and \ref{BoundaryC}\textbf{b} show the inner and outer primary division that results, as well as the boundary classes.  Each inner boundary class has a well defined boundary class as its forward iterate: $\left\{J^*,K, I^*\right\} \rightarrow C_3$, $\left\{J, I, E, D_1, F^*, H^*, G\right\} \rightarrow C_2$, $F \rightarrow C_1$, $G^* \rightarrow C^*$, $D_0 \rightarrow C_0$, $E^* \rightarrow B^*$, $D^* \rightarrow A^*$, $\left\{C_3,C_0\right\} \rightarrow B_1$, $\left\{C^*,C_1,C_2\right\} \rightarrow B_0$, and $\left\{A^*,B^*,B_0,B_1,A\right\} \rightarrow A$.	\\

\textbf{The Secondary Division.}  In order to algorithmically iterate bridge classes forward, particularly for bridges at iterates higher than the maximum iterate of the ETPs, we must first construct the \emph{secondary division} of phase space.  The secondary division is constructed by dividing phase space, as in Fig.~\ref{ConnectionGraphs}\textbf{a}, along the forward iterate of each bridge that has a pseudoneighbor in its interior.  The phase space regions that result from this division are numbered and their connections (at the stable fundamental annulus) are represented in the connection graphs of Fig.~\ref{ConnectionGraphs}\textbf{b}.  The forward iterate of each bridge class will be forced to extend through the appropriate connection graph, and conform to the topology of each numbered region.	\\

	\begin{figure*}[htbp]
		\includegraphics[width = \textwidth]{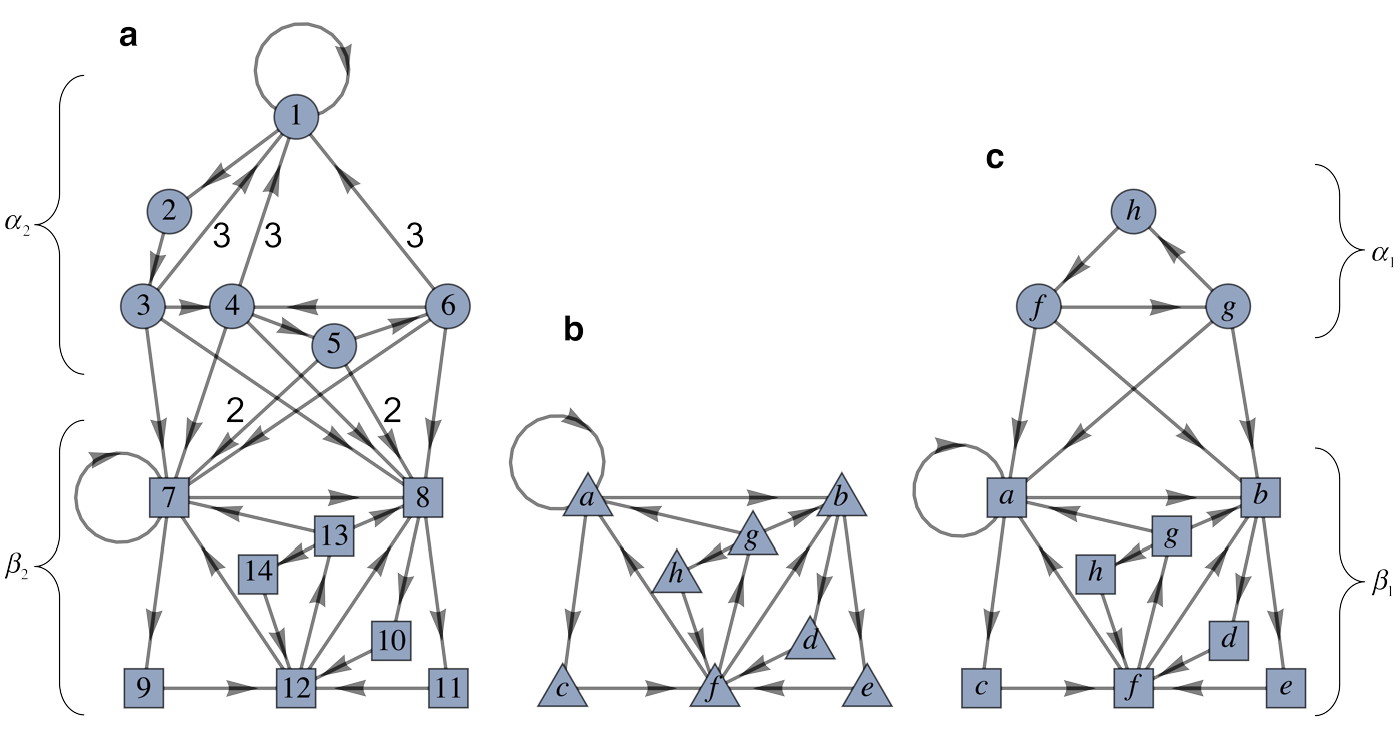}
		\caption{\textbf{Symbolic Dynamics at Iterate Four}.  \textbf{a}, The transition graph representing the dynamics of active, recurrent, 2D bridge classes.  \textbf{b}, The graph representing 1D bridge class dynamics.  \textbf{c}, The ``identified" 1D dynamics graph, which labels the 1D bridge classes of \textbf{b} as being embeddable in 2D bridge classes from either the $\alpha_2$ or $\beta_2$ strongly connected components of \textbf{a}.  If a 1D bridge class can be embedded in 2D bridges of both types then the symbol is split, and has a representative in both $\alpha_1$ and $\beta_1$.}  
		  \label{SymbolicGraphs}
	\end{figure*}

\textbf{Iterating Bridge Classes Forward.}  We can now take a bridge class and iterate all of its boundary classes forward.  We must connect all of these boundary classes together, through the regions given in the connection graph, Fig.~\ref{ConnectionGraphs}\textbf{b}, in the simplest way that is consistent with the topology of each region.  This will in general force the existence of additional boundary classes in the forward iterate.

This forced connection can be done in a unique way, and results in a procedure for iterating any bridge class forward.  At each node of a connection graph, representing a region of the secondary division, our task is to connect the boundary curves together in a way that doesn't intersect any of the 2D unstable boundaries of this region, doesn't include any ``handles" which would change the genus of the overall topology of the unstable manifold, and forces the minimal number of new boundary curves.  Figure~\ref{Forcing} shows an example of this procedure.  In each case of connecting boundary curves through a region, only the concise, minimal connecting surface is topologically unique.  The actual unstable manifold might be more convoluted and topologically complex, which would be revealed by incorporating higher iterate information into our initial ETPs.	\\

\textbf{Inert Boundary Classes.}  The total set of bridge classes closed under iteration can be pruned, while preserving the essential symbolic dynamics, by removing non-recurrent and inert bridge classes.  We can further reduce the size of the symbol set by introducing the concept of an inert \emph{boundary} class, and treat any set of bridge classes whose constituents differ by only inert boundary classes as a single bridge class.  An inert boundary class is defined by the effect its deletion has on the forward iterates of its parent bridge class.  In particular, removing an inert boundary class does not modify the forward iterate of its parent bridge class other than to remove the forward iterate of this inert boundary class.  Additionally, the forward iterate of an inert boundary class must itself be an inert boundary class.  The base case which grounds this recursive definition is given when the bridge class replicates itself under iteration, and the inert boundary class iterates to itself.  For example, the $G$ boundary class in the bridge class $\left[\left[B_1,C_0,F,G\right]\right]$, represented by the symbol 3 in Fig.~\ref{BigSymbolicGraph}\textbf{a}, is inert (red \emph{dashed} circle).  The only effect deleting it would have would be to delete boundary class $C_2$ from the bridge class $\left[\left[B_1,C_1,C_2,D_0,F,G\right]\right]$ (symbol 4 and part of the forward iterate of 3).  $C_2$ in $\left[\left[B_1,C_1,C_2,D_0,F,G\right]\right]$ is itself inert, as we can see by following its forward iterates; first $B_0$ in $\left[\left[B_0,B_0,C_0,C_1,C_2,D_0\right]\right]$ (symbol 5) and then $A$ in the bridge class $\left[\left[A,A,H\right]\right]$ (symbol 6 after removing the inert $A$).  Note that a boundary class is not intrinsically inert, and can only be considered inert with respect to a specific bridge class.  While $C_2$ in $\left[\left[B_1,C_1,C_2,D_0,F,G\right]\right]$ is inert, $C_2$ in $\left[\left[D_1, C_2\right]\right]$ (symbol 7) or $\left[\left[C_2, B_0\right]\right]$ (symbol 8) is not inert.  Deleting $C_2$ from either of these would reduce their forward iterates in Fig.~\ref{BigSymbolicGraph}\textbf{a} to a single trivial cap.		\\

\textbf{4th Iterate ETP and Symbolic Dynamics.}  The analysis in the main text was done using ETP information up to third iterate.  Shown in Fig.~\ref{ETP4num} and Fig.~\ref{ETP4} is the ETP up to iterate four.  Using the HLD analysis presented in the main text and elaborated upon in the online methods, the information embedded in this fourth iterate ETP results in the symbolic graphs of Fig.~\ref{SymbolicGraphs}.  The fourth iterate symbolic entropy associated with Fig.~\ref{SymbolicGraphs}\textbf{a} is $\ln\left(2.1106\right)$.  Again, the graph has two strongly connected components (SCCs), labeled $\alpha_2$ and $\beta_2$, which correspond to a 2D and 1D stretching mechanism respectively.	\\

\textbf{Strong Shift Equivalence.}  As formal Markov shift dynamical systems, the symbolic dynamics embodied in the graphs of Figs.~\ref{BigSymbolicGraph} and \ref{SymbolicGraphs} can be analyzed using a powerful set of tools.  In particular, two graphs can be considered conjugate, i.e. encoding essentially the same dynamics, if they can be connected by a chain of specific vertex merging and splitting operations.\cite{LindMarcus, Kitchens}  This specific relation is also referred to as a strong shift equivalence (SSE).  We have implemented an algorithm which searches for such a chain connecting two given transition matrices.  First, for both transition matrices, we generate every matrix less than a given dimension that is reachable through no more than a given number of such merging and splitting moves.  As this process unfolds, we periodically prune each set such that no two matrices are simply related by index permutations.  We then compare the matrices in both accumulated sets.  If there is a permutation-related pair, then the two original transition matrices are conjugate.  If not, then we expand the length of allowable chains and repeat the procedure.  While this algorithm can find an SSE, it cannot prove the absence of such an equivalence, as often the connecting chains of matrices are either quite long, or require that at least one of their members be a matrix of dimension much higher than the original transition matrices.  

Determination of strong shift equivalence has allowed us to identify the duality between the forward time 2D stretching mechanism and the backward time 1D stretching mechanism.  Additionally, the merging operation provides a convenient way of identifying bridge classes (vertices) that can be merged, and therefore should be considered equivalent.  Indeed, this serves as an algorithmic way of checking for the existence of inert boundary classes, or whether unnecessary bridge classes were originally used (say due to superfluous obstruction rings).\\

\normalsize
\textbf{References}\small
\bibliography{HillsHLD3DBib}

\normalsize
\textbf{Acknowledgements}\\
We would like to acknowledge Haik Stepanian for preliminary numerical work on computing stable and unstable manifolds.  We would also like to thank Roy Goodman, Jacek Wr\'{o}bel, Hector Lomeli, and Jason Mireles James for extensive discussions on computing these manifolds.  This work was supported in part by the US DOD, ARO grant W911NF-14-1-0359 under subcontract C00045065-4.

\textbf{Author Contributions}\\
SAS carried out the HLD analysis of the scattering data and did the bulk of the manuscript writing and preparation.  JA performed the numerical computations of the scattering data.  ER and SS provided the independent topological entropy computation following Ref.~\cite{HuntOtt}.  KAM was involved in all aspects of this work.  All authors helped revise and edit the text.

\end{document}